\documentclass[10pt,twocolumn,letterpaper]{article}

 \usepackage[pagenumbers]{cvpr} 

\usepackage{algorithm}
\usepackage{algorithmicx}

\usepackage{algpseudocode}

\definecolor{cvprblue}{rgb}{0.21,0.49,0.74}
\usepackage[pagebackref,breaklinks,colorlinks,allcolors=cvprblue]{hyperref}

\def\paperID{12269} 
\def\confName{CVPR}
\def\confYear{2025}

\title{Active Sampling and Gaussian Reconstruction for \\ Radio Frequency Radiance Field}

\author{Chi-Shiang Gau\\
UC San Diego\\
{\tt\small gau@ucsd.edu}
\and
Xingyu Chen\\
UC San Diego\\
{\tt\small xic063@ucsd.edu}
\and
Tara Javidi\\
UC San Diego\\
{\tt\small tjavidi@ucsd.edu }
\and
Xinyu Zhang\\
UC San Diego\\
{\tt\small xyzhang@ucsd.edu }
}

\begin{document}
\maketitle
\vspace{10pt}

\begin{abstract}
Radio-Frequency (RF) Radiance Field reconstruction is akin to view synthesis in computer vision, but entails more challenges due to the complex interactions between RF signals and objects, including reflections and diffraction. These interactions are difficult to model accurately, especially when the shapes and materials of the objects are unknown. A recent neural network-based method called NeRF2 proposed a differentiable, large-parameter model for reconstructing and predicting RF signals in a given environment. However, NeRF2's dependence on an excessive number of samples and its compute-intensive ``training'' process for each environment limits its practical applicability, particularly in dynamic or quasi-static settings. Additionally, NeRF2 lacks an uncertainty model for its predictions. Inspired by 3DGS, we propose a training-free method that replaces NeRF2's neural model with a Gaussian process using an adaptive spatial-temporal kernel. This kernel has a tunable hyperparameter (locality parameter in space and time), which captures side information about the scenario. Specifically, we demonstrate that:
{\textbf{Gaussian models require significantly fewer measurements}} than neural models. Furthermore, incorporating an uncertainty model, posterior (co-)variance points to maximally-informative samples in space/time. This results in {\textbf{further reduction in the required number of samples via active sampling,}} and {\textbf{the first efficient method to predict RF signals in dynamic environments.}}
\vspace{-10pt}
\end{abstract}
    
\section{Introduction}
\label{sec:intro}

The modeling of electromagnetic fields in complex indoor and outdoor environments, particularly at radio frequency (RF) bands, has become increasingly critical for emerging applications in network planning and wireless communications \cite{hu2023muse,zhao2022high,vakalis2019imaging,woodford2022mosaic}. RF signal propagation is fundamentally shaped by material properties and wave phenomena including reflection, refraction, and diffraction from object geometries in the scene. Similar to view synthesis for cameras, RF Radiance Field modeling aims to determine the total received power at any given spatial position.


A recent work, called NeRF2 \cite{zhao2023nerf2} has explored adapting neural-based novel view synthesis (NVS) techniques, such as Neural Radiance Fields (NeRF) \cite{mildenhall2021nerf}, from the visual to the radio domain, yielding promising results. However, two significant challenges impede the practical deployment of these methods. First, due to the limited aperture size of radio antennas compared to optical sensors, these approaches require an exceptionally high sampling density—approximately 200 samples per square foot, in contrast to the dozens of images per scene typically needed for optical NVS. Second, radio systems demand real-time performance and often operate under computational constraints, making iterative gradient-descent optimization methods like neural networks or Gaussian Splatting \cite{kerbl20233d} impractical for real-time radio frequency reconstruction.

To address these limitations, we propose a Gaussian reconstruction method that leverages Gaussian processes to model RF Radiance Fields. Our key insight is to represent these fields as spatially dense Gaussian fields, populated with virtual signal sources, modeled as Gaussian random variables. This allows us to construct RF Radiance Fields by resolving field uncertainties through sparse RF samples. Additionally, our method allows us to provide an uncertainty model. In other words, for any selected target position, we not only predict the RF Radiance Field, and hence received signal power, but also provide the variance of the predicted value.

The practical contribution of our work is to incorporate active sampling by taking measurements at the positions with the highest uncertainty, which significantly reduces the required sampling density. Moreover, with this approach, our method can adapt to quasi-dynamic scene changes by collecting only a few measurements and comparing them with the past RF Radiance Field. When changes are detected, our method guides the agent in identifying which additional measurements are needed, allowing the RF Radiance Field to be updated accordingly. This capability distinguishes our method from existing approaches and makes it more suitable for real-world applications.

We evaluate our method in both simulated and real environments, demonstrating superior performance in terms of computational and sampling efficiency, while maintaining high reconstruction accuracy. As a result, our proposed method is particularly advantageous when the number of available samples is limited.


Our contributions can be summarized as follows:

\begin{itemize}

    \item \textbf{No pre-training required}: Our proposed method does not require any pre-training to make predictions for any selected target positions, saving us several hours compared to NeRF2.

    \item \textbf{An uncertainty model is provided}: We provide an uncertainty model for any given position in the RF Radiance Field, allowing users to determine if additional observations are needed to reduce uncertainty in specific areas—something that NeRF2 does not offer.

    \item  \textbf{Reduction in the number of samples}: With the help of our uncertainty model, highly informative samples are collected first, reducing the number of samples needed to achieve performance comparable to NeRF2. Our experiments demonstrate a reduction of $30\%$ to $60\%$ in the number of samples compared to NeRF2.

    \item  \textbf{Fast adaptation to quasi-dynamic scene changes}: Our proposed method efficiently adapts to changes in the scene by focusing on learning the variations to reconstruct the new RF Radiance Field.

\end{itemize}
\section{Related Works}


Neural representations for scene understanding have revolutionized computer vision and graphics \cite{mildenhall2020nerf, park2019deepsdf, sitzmann2020implicit, tewari2020state, yariv2020multiview, chan2021pifu,xie2021neural}, particularly since the seminal work of Neural Radiance Fields (NeRF) \cite{mildenhall2021nerf}. NeRF demonstrated that complex 3D scenes can be represented through implicit neural networks that encode both spatial geometry (via volumetric density) and view-dependent appearance (via directional radiance). This enables photorealistic novel view synthesis without explicit 3D modeling.
Recent work such as NeRF2~\cite{zhao2023nerf2} has extended neural scene representations beyond visible light to RF domains.  Subsequent approaches have explored different representation strategies. For example,  WiNeRT~\cite{orekondy2023winert} bridges neural and physics-based methods through learnable material-specific reflection parameters, while RFCanvas~\cite{chen2024rfcanvas} introduces a fully explicit representation that can adapt to RF field dynamics. However, these methods are fundamentally limited by their reliance on optimization-based training from dense RF measurements, making them impractical for active learning scenarios where sample efficiency is crucial.

In contrast to these approaches, we propose modeling RF radiance fields using Gaussian random process, eliminating the need for optimization-based training. Our method enables active learning by providing precise control over sampling strategies, making it particularly suitable for real-world applications where data collection is costly or time-constrained.
\section{Preliminary}
In this section, we present an overview of the system model and the fundamental properties of Gaussian random processes.

\subsection{Notations}

We use boldface letters to represent random variables and random vectors. $\Vec{(.)}$ denotes a vector, while capital letters represent matrices. $(\cdot)^T$ denotes the transpose operation. $\mathbb{E}$ is the expectation operator.

\subsection{System Model and Ray Structure}
\begin{figure}[h]
\centering
\includegraphics[width=0.95\linewidth]{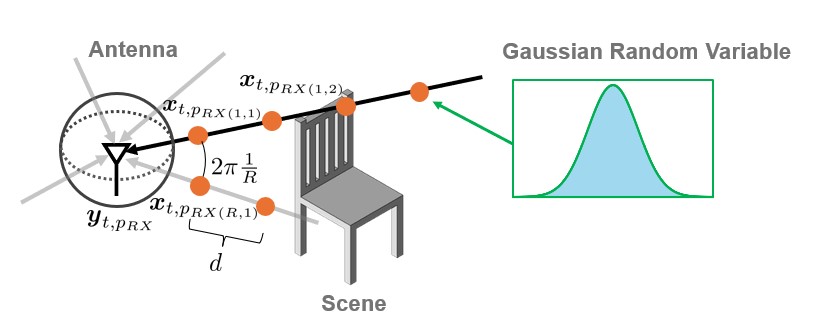}
\caption{The Radiance field in the RF domain, measured by an omnidirectional antenna, has a simple (radial) ray structure; furthermore, the RF signal is impacted by significant path loss, resulting in a smaller "cross-talk" and a simpler inverse problem. }
\label{fig: ray structure}
\end{figure}

We assume that the scene remains static during a given time slot. We model each position in the room as an omnidirectional virtual signal source. Specifically, for time slot $t$, the virtual signal at position $p_i$, is denoted as $\boldsymbol{x}_{t,p_i}$. Furthermore, the received signal power at any position is modeled as a sum of $R$ different rays from different angles. The received signal from each ray is modeled as a linear combination of the $N$ virtual signal source sampled on the ray, as shown in Fig.~\ref{fig: ray structure}.

For time slot $t$ the received signal can be written as:
 \begin{align}
    \boldsymbol{y}_{t,p_{RX}}
    & \triangleq \sum_{r=1}^R \sum_{n=1}^N \alpha_{p_{RX(r,n)}} \cdot \boldsymbol{x}_{t,p_{RX(r,n)}} \label{eq:received signal model}
\end{align} 

\noindent where $\boldsymbol{y}_{t,p_{RX}}$ is the received signal power at position $p_{RX}$. The terms $\boldsymbol{x}_{t,p_{RX(r,n)}}$  represent the virtual signal source at position ${p_{RX(r,n)}}$. Additionally, $\alpha_{p_{RX(r,n)}}$ denotes the attenuation from the virtual signal source at position ${p_{RX(r,n)}}$ to the receiver at position $p_{RX}$. In this work, the receiver is assumed to have a single omnidirectional antenna. As such, the attenuation term simply represents propagation loss. However, if a directional antenna or antenna array is used at the receiver, both the antenna factor and the array factor would contribute to the attenuation term, $\alpha_{p_{RX(r,n)}}$.

The notation $RX(r,n)$ refers to the position of the $n-$th sample on the $r-$th ray for the receiver placed at $p_{RX}$. Specifically, there are $R\times N$ virtual signal sources that contribute to the received signal at $p_{RX}$ and their corresponding positions are given by:
\begin{align}
    &p_{RX(r,n)} \triangleq p_{RX} + n \cdot d \cdot 2\pi\frac{r}{R}, \nonumber \\
    &\text{for} \ r=1,2,..., R \ \text{and} \ n=1,2,..., N
\end{align}

\noindent where $d$ is the sampling resolution on the ray. The propagation loss for the virtual signal source $p_{RX(r,n)}$ toward receiver position $p_{RX}$ is:
\begin{align}
    \alpha_{p_{RX(r,n)}} \triangleq \frac{\beta}{|p_{RX(r,n)} - p_{RX}|} = \frac{\beta}{n \cdot d}
\end{align}

\noindent where $\beta$ is the parameter for the transmission medium. In this work, we adopt a simple path loss model as in \cite{zhao2023nerf2}. A more accurate model could be used if additional information, such as the transmission frequency, is available.

For notation simplicity, we rewrite the received signal model for time slot $t$ (\ref{eq:received signal model}) into vector form:
\begin{align}
    \boldsymbol{y}_{t,p_{RX}} \triangleq \vec{\alpha}_{p_{RX}}^T \cdot \vec{\boldsymbol{x}}_{t,p_{RX}} \label{eq: observation vector form}
\end{align}

Notice that, although the receiver is located within the room, some virtual signal sources may be located outside. This approach accounts for situations where the receiver is placed at the room's boundary or in a corner.

\subsection{Gaussian Random Field}
For a given time slot $t$, we model the RF Radiance Field as a Gaussian random process, with the virtual signal sources—components of the RF Radiance Field—modeled as Gaussian random variables. A Gaussian process is defined by its mean and covariance  (also known as the kernel):

\begin{itemize}
\item \textbf{Mean}: 
For any position $p_i$, the virtual signal is modeled as a zero-mean Gaussian random variable. 
\begin{align}
    \mathbb{E} \left[ \boldsymbol{x}_{t, p_i} \right] = 0
\end{align}

\item \textbf{Covariance}: The covariance between each pair of virtual signal sources is given by:
\begin{align}
Cov(\boldsymbol{x}_{t,p_i},\boldsymbol{x}_{t,p_j}) = \alpha^2 \exp{\left(-\frac{1}{2l^2}||p_i - p_j||^2\right)} \label{eq: RBF kernel}
\end{align}
\end{itemize}

The value of $a^2$ represents the variance of the virtual signal sources, which are selected based on the true transmission power of the device within the room.  The length-scale parameter $l$, controls the rate of variation of the signal; smaller values of 
$l$ indicate a sharper change in the function. In a later section, we will demonstrate how to estimate this parameter. Note that (\ref{eq: RBF kernel}) represents the Radial Basis Function (RBF) kernel.  
For information on other types of kernel functions and guidance on selecting the appropriate kernel, we refer the reader to \cite{williams2006gaussian}.

With the above definitions, $\vec{\boldsymbol{x}}_{t,p_{RX}}$, the collection of virtual signal sources contributing to $\boldsymbol{y}_{t,p_{RX}}$ is a Gaussian random vector:
\begin{align}
\vec{\boldsymbol{x}}_{t,p_{RX}} \sim N(\vec{0}, \Sigma_{p_{RX}})
\end{align}
Here, $\Sigma_{p_{RX}}$ is the covariance matrix where the $(i,j)-$th component represents the covariance between the $i-th$ and $j-th$ components of $\vec{\boldsymbol{x}}_{t,p_{RX}}$.

Also, $\boldsymbol{y}_{t,p_{RX}}$, a linear combination of Gaussian random variables.  (\ref{eq: observation vector form}) is a Gaussian random variable with the following distribution:
\begin{align}
    \boldsymbol{y}_{t,p_{RX}} \sim N(0, \vec{\alpha}_{p_{RX}}^T\Sigma_{p_{RX}}\vec{\alpha}_{p_{RX}})  \label{eq:one observation}
\end{align}

\subsection{Conventional Gaussian Prediction}
Next, we describe how to predict the received signal power $\boldsymbol{y}_{t,p_{Target}}$ at a given target position $p_{Target}$ using $M$ observations $\boldsymbol{y}_{t,p_{RX_1}}, \boldsymbol{y}_{t,p_{RX_2}}, ..., \boldsymbol{y}_{t,p_{RX_M}}$, within the same time slot $t$. Since the observation is a Gaussian random variable, the collection of $M$ observations can be represented as a Gaussian random vector:
\begin{align}
\vec{\boldsymbol{y}}_{t,p_{RX_{1:M}}} \triangleq 
\begin{bmatrix}
\boldsymbol{y}_{t,p_{RX_1}} \\
\boldsymbol{y}_{t,p_{RX_2}} \\
\vdots \\
\boldsymbol{y}_{t,p_{RX_M}}
\end{bmatrix} = A \cdot \vec{\boldsymbol{x}}_{t,p_{RX_{1:M}}} \label{eq: past observations}
\end{align}

\noindent where $\vec{\boldsymbol{x}}_{t,p_{RX_{1:M}}}$ represents the collection of virtual signal sources contributing to the $M$ observations for time slot $t$: 
\begin{align}
\vec{\boldsymbol{x}}_{t,p_{RX_{1:M}}} \triangleq 
\begin{bmatrix}
\boldsymbol{x}_{t,p_{RX_1}} \\
\boldsymbol{x}_{t,p_{RX_2}} \\
\vdots \\
\boldsymbol{x}_{t,p_{RX_M}}
\end{bmatrix}
\end{align}

The matrix $A$ is the linear combination matrix, given by:
\begin{align}
    A \triangleq I_M \otimes \vec{\alpha}_{p_{RX}}^T
\end{align}
where $I_M$ is the $M \times M$ identity matrix and $\otimes$ represents the Kronecker product.

Next, by stacking the the received signal power $\boldsymbol{y}_{t,p_{Target}}$ for the target position $p_{Target}$ into (\ref{eq: past observations}), we obtain a new Gaussian random vector:
\begin{align}
\left[\begin{array}{c} 
	\vec{\boldsymbol{y}}_{t,p_{RX_{1:M}}} \\ 
	\hline 
	\boldsymbol{y}_{t,p_{Target}}   
\end{array}\right] 
= 
\left[\begin{array}{c|c} 
	A & \vec{0} \\ 
	\hline 
	\vec{0} & \vec{\alpha}_{p_{RX}}^T  
\end{array}\right] 
\cdot 
\left[\begin{array}{c} 
	\vec{\boldsymbol{x}}_{t,p_{RX_{1:M}}} \\ 
	\hline 
	\boldsymbol{x}_{t,p_{Target}}   
\end{array}\right] \label{eq: stack target}
\end{align}

The mean and covariance matrix can be computed using (\ref{eq:one observation}). We denote this distribution as:
\begin{align}
&\left[\begin{array}{c} 
	\vec{\boldsymbol{y}}_{t,p_{RX_{1:M}}} \\ 
	\hline 
	\boldsymbol{y}_{t,p_{Target}}   
\end{array}\right] 
\sim \nonumber
\\
& N( 
\left[\begin{array}{c} 
	\vec{\mu}_{t,Past} \\ 
	\hline 
	\mu_{t,Target}   
\end{array}\right] , 
\left[\begin{array}{c|c} 
 \Sigma_{t,Past} & \Sigma_{t,Past,Target} \\ 
 \hline 
 \Sigma_{t,Target,Past} & \Sigma_{t,Target}  
\end{array}\right] ) \label{eq: distribution of the new vector with target}
\end{align}
Note that from (\ref{eq:one observation}) and (\ref{eq: past observations}), $\vec{\mu}_{t,Past}$  is a zero vector and $\mu_{t,Target}$ is a zero scalar, respectively.

With (\ref{eq: distribution of the new vector with target}), the predicted value $\boldsymbol{y}_{t,p_{Target}}$ for the target position $p_{Target}$ can be calculated using the conditional probability formula:
\begin{align}
\hat{\boldsymbol{y}}_{t,p_{Target}} \triangleq \boldsymbol{y}_{t,p_{Target}} | \{ \vec{\boldsymbol{y}}_{t,p_{RX_{1:M}}} =  \vec{y}_{t,p_{RX_{1:M}}} \}
\end{align}

\begin{itemize}
\item \textbf{Prediction Mean}: 
\begin{align} 
&\text{mean}(\hat{\boldsymbol{y}}_{t,p_{Target}}) = \nonumber \\ &\Sigma_{t,Target,Past}
\Sigma^{-1}_{t,Past}(\vec{y}_{t,p_{RX_{1:M}}} - \vec{\mu}_{t,Past}) \label{eq: prediction mean}
\end{align}
\item \textbf{Prediction Variance}: 
\begin{align}
&\text{var}(\hat{\boldsymbol{y}}_{t,p_{Target}}) = \nonumber
\\
&\Sigma_{t,Target} - \Sigma_{t,Target,Past}
\Sigma^{-1}_{t,Past} \Sigma_{t,Past,Target}  \label{eq: prediction var}
\end{align}

\end{itemize}

Notice that in (\ref{eq: prediction var}), the prediction variance at the selected target position depends only on the target position and the locations of all previously sampled observations.

\section{Methods}


Our approach consists of three main parts: Local Kernel Estimation, Active Sampling, and Quasi-Dynamic Reconstruction. First, we use the given observations to reconstruct the RF Radiance Field with a more precise uncertainty model compared to equation (\ref{eq: prediction var}). Next, leveraging this uncertainty model, we show how observations can be collected more efficiently. Finally, we demonstrate that when the RF Radiance Field changes, only a few new samples are needed to reconstruct the updated field.

\subsection{Local Kernel Estimation}
The computational bottleneck for conventional Gaussian prediction arises from the matrix inversion of $\Sigma_{t,Past}$ in (\ref{eq: prediction mean}) and (\ref{eq: prediction var}). The size of $\Sigma_{t,Past}$ for $M$ observations is $(MRN)\times (MRN)$. As a result, the computational complexity is $\mathcal{O}((MRN)^3)$. This complexity is particularly high in large-scale radiance fields which entail a large number of observations.

Moreover, in a given scene, certain areas may consist of empty space, where virtual signal sources are expected to be highly correlated. In contrast, other areas may be densely populated with objects such as tables, chairs, and other furniture. In these crowded regions, the correlation between virtual signal sources will differ from that in the empty spaces. As a result, the uncertainty model should be adapted based on the characteristics of the observations. Specifically, we assign low uncertainty to smoother areas and higher uncertainty to regions with sharper variations.

To address the two issues mentioned above, we propose a \textit{local kernel estimation} strategy that tackles both the computational complexity and the correlation locality problems. For a selected target position $p_{Target}$, we use only the past observations sampled close to $p_{Target}$ to perform the local kernel estimation and predict $\boldsymbol{y}_{t,p_{Target}}$. Specifically, observation $\boldsymbol{y}_{t,p_{RX_i}}$ will be used for the local kernel estimation for position $p_{Target}$, if the position of the observation, $p_{RX_i}$, is sufficiently close $p_{Target}$, that is:
\begin{align}
    \{ p_{RX_i} \mid \| p_{RX_i} - p_{Target} \| \leq L \}  \label{locality}
\end{align}

Here, $L$ is the locality parameter, which can be set based on the complexity of the scene structure. In this work, we present the concept of local kernel estimation, 
leaving the determination of the optimal value for $L$ for future work.

Once the observations for the selected target are obtained, we need to estimate the kernel’s length-scale parameter $l$, as shown in (\ref{eq: RBF kernel}). However, since we are now performing local kernel estimation, this length-scale parameter $l$ should depend on the position of the target. Therefore, we modify (\ref{eq: RBF kernel}) as follows:
\begin{align}
    Cov(\boldsymbol{x}_{t,p_i},\boldsymbol{x}_{t,p_j}) = \alpha^2 \exp{\left(-\frac{1}{2l_{t,p_{Target}}^2}||p_i - p_j||^2\right)} \label{eq: local RBF kernel}
\end{align}

Next, we perform maximum likelihood estimation to estimate the local length-scale parameter $l_{t,p_{Target}}$ in (\ref{eq: local RBF kernel}):
\begin{align}
    \hat{l}_{t,p_{Target}} = \arg \max_l \frac{1}{(2\pi)^\frac{M_{local
    }}{2}|\Sigma_{t,l}|^\frac{1}{2}} e^{-\frac{1}{2}\vec{\boldsymbol{y}}_{t,local}^T\Sigma_{t,l}^{-1}\vec{\boldsymbol{y}}_{t,local}} \label{eq: maximum likelihood}
\end{align} 

\noindent where $\vec{\boldsymbol{y}}_{t,local}$ is a subset of $\vec{\boldsymbol{y}}_{t,p_{RX_{1:M}}}$ that satisfies the condition in (\ref{locality}) and $M_{local
    }$ denotes the number of elements of $\vec{\boldsymbol{y}}_{t,local}$. Note that the local length-scale parameter $l_{t,p_{Target}}$ indicates which areas are smooth and which exhibit sharper spatial variations. After estimating the local kernel, we can use equations (\ref{eq: prediction mean}) and (\ref{eq: prediction var}) to calculate the prediction mean and variance of $\boldsymbol{y}_{t,p_{Target}}$.

Since only $M_{local}$ observations are used for the prediction, the computation complexity is reduced from $\mathcal{O}((MRN)^3)$ to $\mathcal{O}((M_{loacl}RN)^3)$. If the observations are uniformly distributed within the room, then the ratio $\frac{M_{local}}{M}$ is approximately$\frac{2\pi L^2}{\text{size of the scene}}$.

\begin{figure}[h]
\centering
\includegraphics[width=\linewidth,height=3cm]{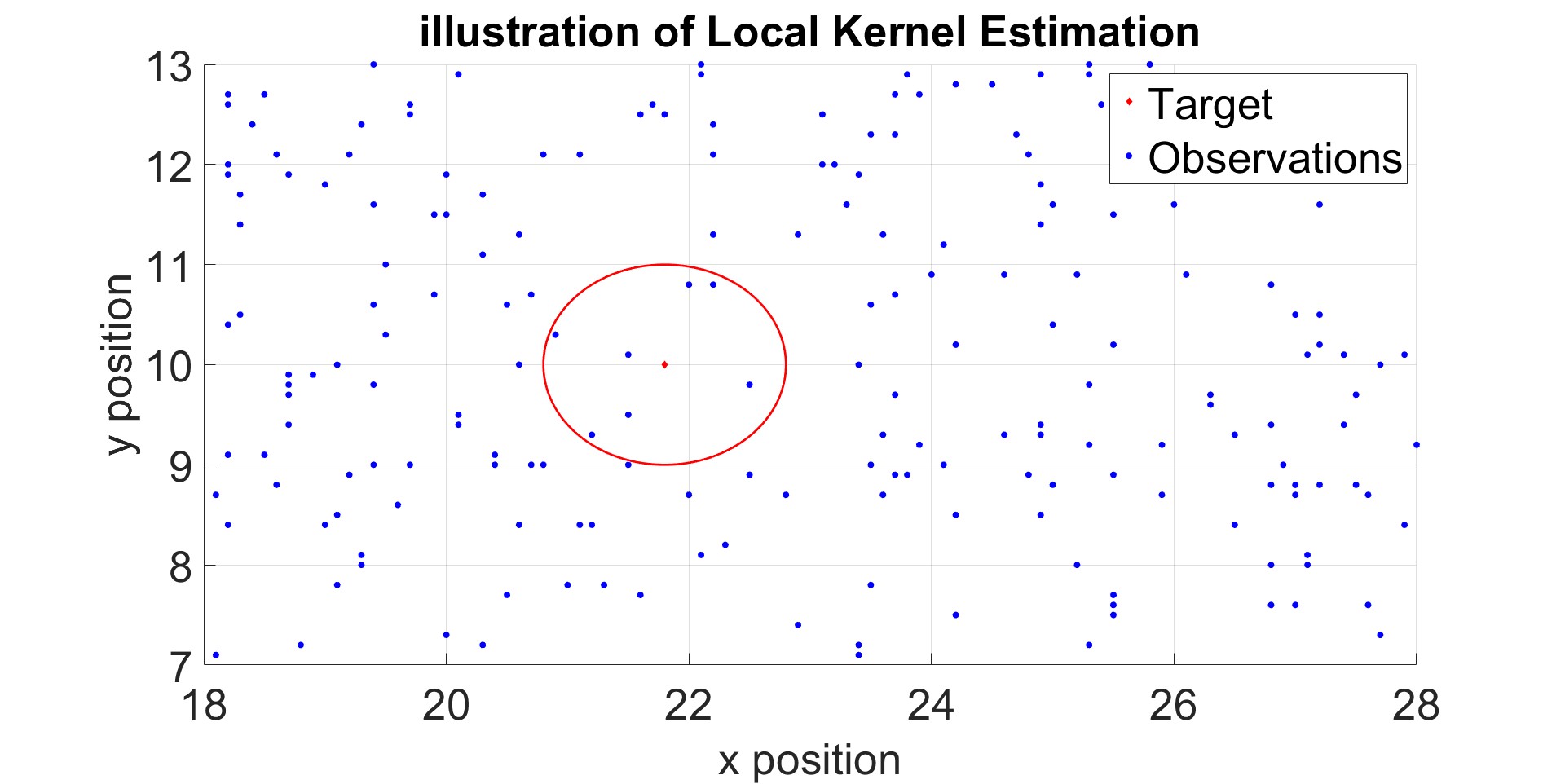}
\caption{illustration of the Local Kernel Estimation}
\label{fig: local kernel estimation}
\end{figure}

\begin{figure}
\begin{minipage}[t]{0.49\columnwidth}
  \includegraphics[width=\linewidth]{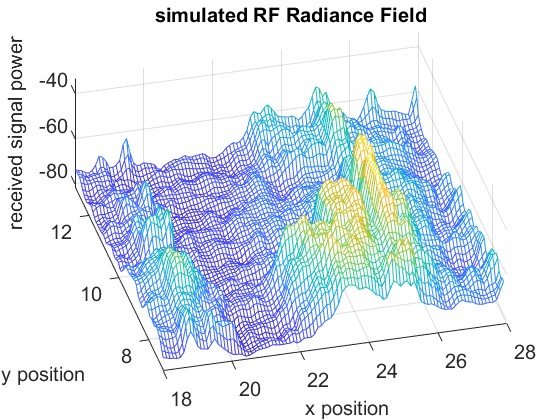}
\caption{Simulated RF Radiance Field}
\label{fig: local_kernel_estimation_function}
\end{minipage}\hfill 
\begin{minipage}[t]{0.49\columnwidth}
\includegraphics[width=\linewidth]{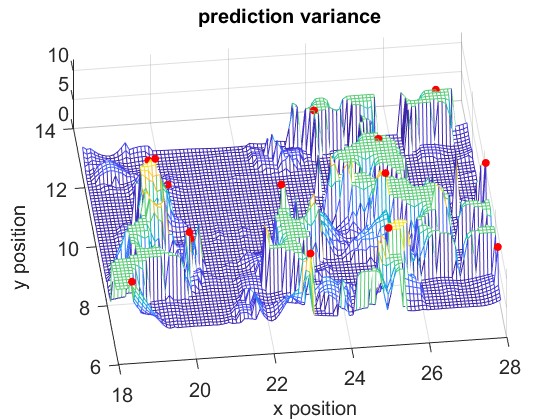}
\caption{Prediction variance and selected next samples}
\label{fig: prediction var}
\end{minipage}

\begin{minipage}[t]{0.49\columnwidth}
\includegraphics[width=\linewidth]{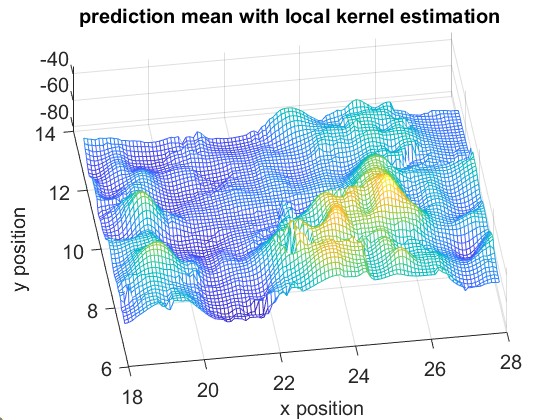}
\caption{Prediction mean with local kernel estimation}
\label{fig: prediction mean with LocalKerEst}
\end{minipage}\hfill 
\begin{minipage}[t]{0.49\columnwidth}
\includegraphics[width=\linewidth]{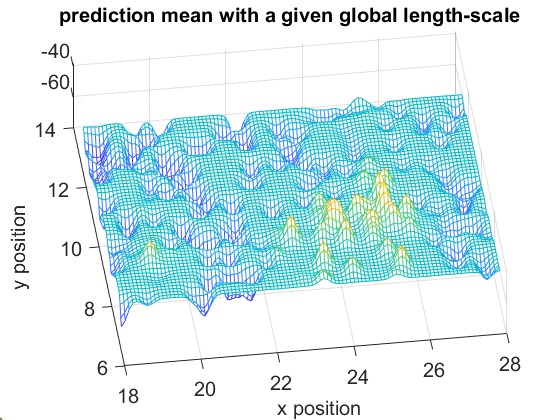}
\caption{Prediction mean without local kernel estimation}
\label{fig: prediction mean without LocalKerEst}
\end{minipage}
\end{figure}

To illustrate the concept of local kernel estimation, we collect 200 observations from simulated data generated by the NeRF2 pre-trained model, which simulates a room measuring 
$10m \times 6m$, as shown in Fig.\ref{fig: local_kernel_estimation_function}. The locality parameter is set to $L = 1m$. As shown in the red circle of Fig.~\ref{fig: local kernel estimation}, 
only 7 observations are needed to predict the received signal at the selected target position.

Fig.~\ref{fig: prediction mean with LocalKerEst} and Fig.~\ref{fig: prediction mean without LocalKerEst} 
show the reconstruction results with and without our proposed local kernel estimation. Compared to the ground truth in Fig.~\ref{fig: local_kernel_estimation_function}, it is clear that the local kernel estimation helps capture the shape details of the RF Radiance Field, leading to a more accurate reconstruction.

\subsection{Active Sampling}
As mentioned earlier, collecting new observations can be costly, and some may offer limited or negligible information. In such cases, adding these samples to the reconstruction may not significantly improve the model’s accuracy. To address this, we propose an active sampling strategy to focus on collecting more informative observations.

Our approach starts with an initial prediction of the RF Radiance Field using a small number of random observations. We then choose the next observation position based on the highest predicted variance, as given by (\ref{eq: prediction var}). Fig.~\ref{fig: prediction var} shows an example of the prediction variance of the received signal power across different positions. After a few initial observations, only certain areas of the scene exhibit high uncertainty (i.e., large variance). Thus, further observations are focused on these high-variance regions to improve the model’s accuracy most efficiently. Note that the high-variance regions in Fig.~\ref{fig: prediction var} match the areas of sharp change in Fig.~\ref{fig: local_kernel_estimation_function}, which are exactly the regions that require more samples to learn. The proposed method for static RF Radiance Field is summarized in Alg.~\ref{alg: static}. 

In this work, instead of only selecting the largest variance position each time, we divide the scene into smaller sections and select the position with the highest variance within each section as the next observation point. This approach helps avoid repeatedly computing the matrix inverse in (\ref{eq: prediction mean}) and (\ref{eq: prediction var}) by adding only one new observation at a time. Note that some sections may be skipped if the predicted variances of all positions within a section are below a certain threshold. 

In summary, Local Kernel Estimation helps us learn the shape and smoothness of different regions in the RF Radiance Field. Prediction variance shows which areas need more samples and which have little uncertainty. Using this variance, the active sampler efficiently avoids oversampling and undersampling.

\begin{algorithm}[ht]
\caption{Proposed method for static RF Radiance Field} \label{alg: static}
\textbf{Input:}
\begin{itemize}
    \item $M$: Initial number of observations
    \item targets: Set of target positions
\end{itemize}

\textbf{Output:} Predicted mean and variance for each position

\textbf{Initialize:} Collect $M$ initial observations 

\textbf{For each target position} in targets:
\begin{algorithmic}
\State Step 1: Find local samples close to $P_{Target}$ from the initial observations\;
\State Step 2: Estimate the local kernel based on the selected local samples\;
\State Step 3: Calculate the prediction mean and variance using the estimated local kernel and the local samples\;

\If{Active Sampling}
    \State Collect new samples at the position with the largest variance (within each section or for the entire scene), then return to Step 1.
\EndIf
\end{algorithmic}
\end{algorithm}

\begin{figure*}[ht!]
\centering
\begin{subfigure}{0.245\textwidth}
\includegraphics[width=0.98\linewidth]{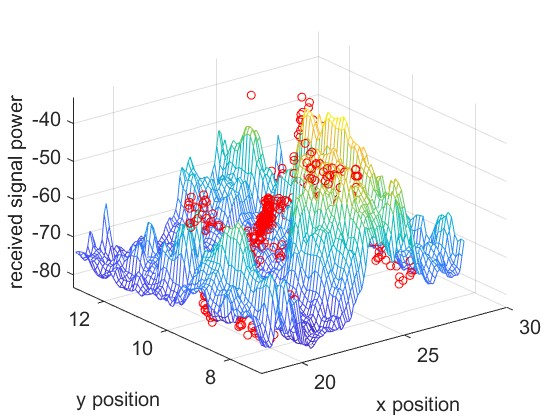} 
\caption{Scenario 1}
\end{subfigure}
\begin{subfigure}{0.245\textwidth}
\includegraphics[width=0.98\linewidth]{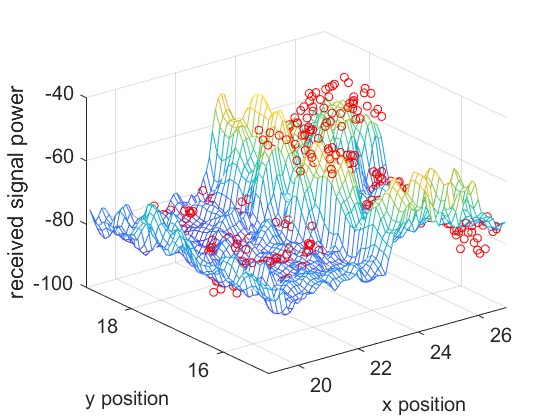}
\caption{Scenario 2}
\end{subfigure}
\begin{subfigure}{0.245\textwidth}
\includegraphics[width=0.98\linewidth]{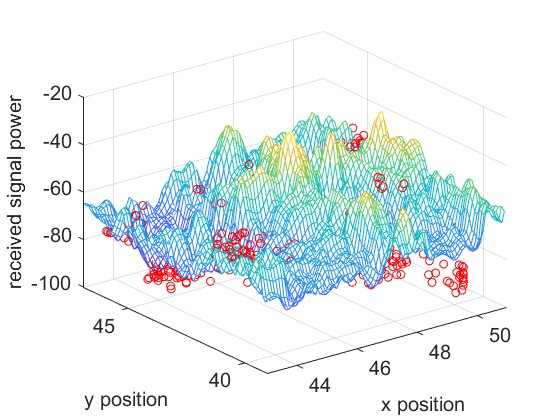}
\caption{Scenario 3}
\end{subfigure}
\begin{subfigure}{0.245\textwidth}
\includegraphics[width=0.98\linewidth]{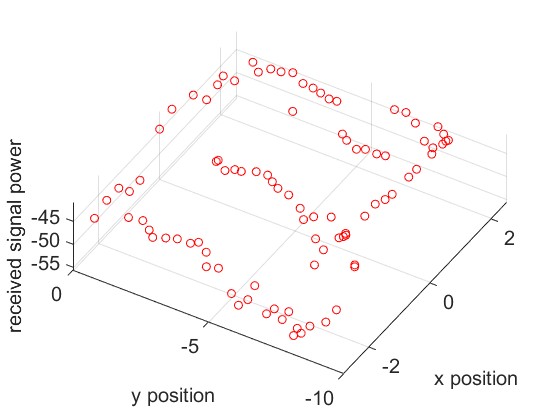}
\caption{Scenario 4}

\end{subfigure}

\caption{Simulated and Real data for various scenarios}
\label{fig: all NeRF2 scenario}
\end{figure*}

\begin{figure*}[ht!]
\centering
\begin{subfigure}{0.33\textwidth}
\includegraphics[width=0.9\linewidth,height=3.5cm]{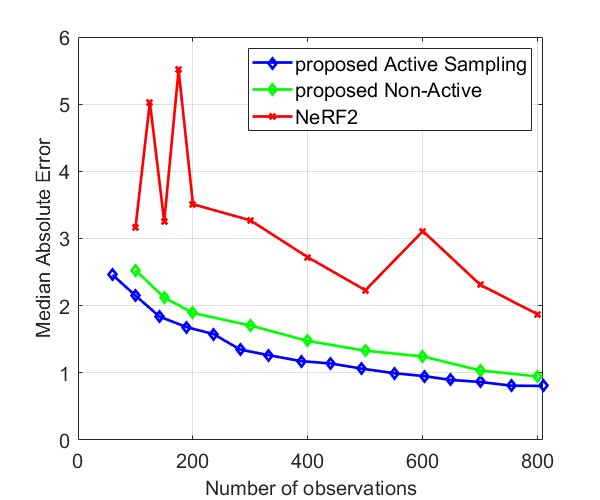} 
\caption{Median AE for Scenario 1}
\end{subfigure}
\begin{subfigure}{0.33\textwidth}
\includegraphics[width=0.9\linewidth,height=3.5cm]{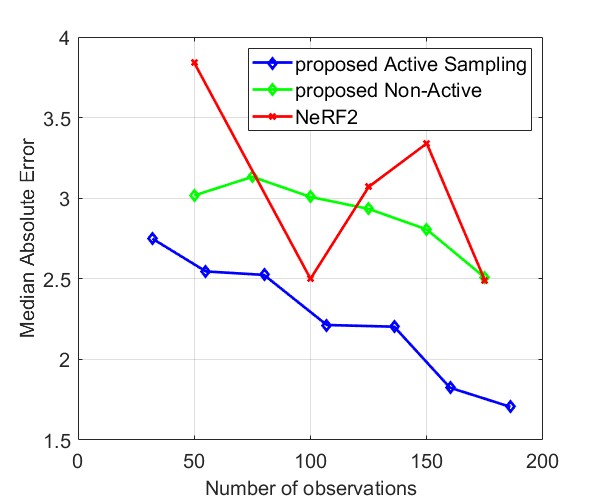}
\caption{Median AE for Scenario 2}
\end{subfigure}
\begin{subfigure}{0.33\textwidth}
\includegraphics[width=0.9\linewidth,height=3.5cm]{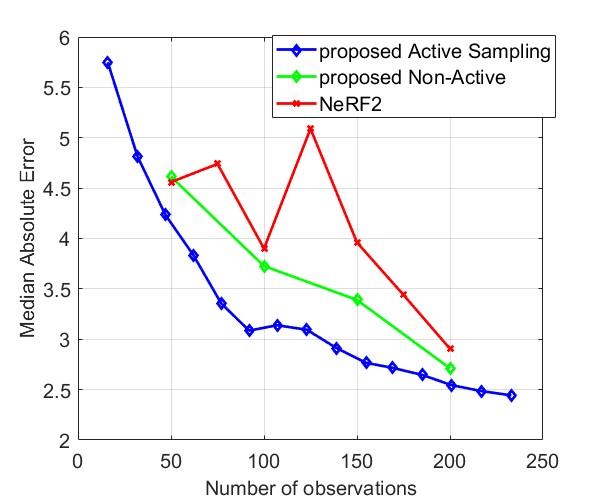}
\caption{Median AE for Scenario 3}
\end{subfigure}

\begin{subfigure}{0.33\textwidth}
\includegraphics[width=0.9\linewidth,height=3.5cm]{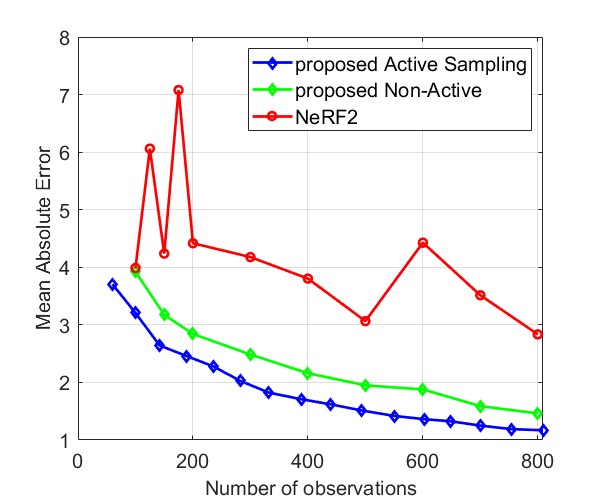} 
\caption{MAE for Scenario 1}
\end{subfigure}
\begin{subfigure}{0.33\textwidth}
\includegraphics[width=0.9\linewidth,height=3.5cm]{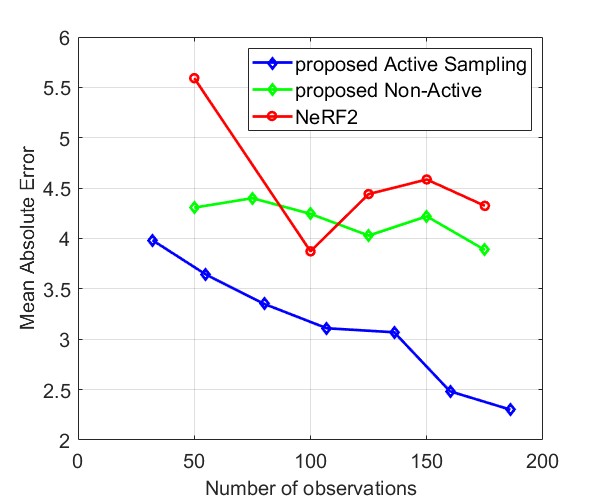}
\caption{MAE for Scenario 2}
\end{subfigure}
\begin{subfigure}{0.33\textwidth}
\includegraphics[width=0.9\linewidth,height=3.5cm]{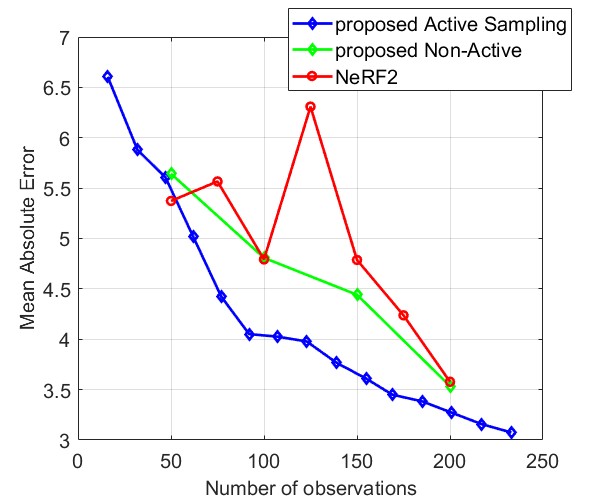}
\caption{MAE for Scenario 3}
\end{subfigure}

\caption{Performance comparison for various scenarios using simulated data}
\label{fig: Median AE simulated data}

\end{figure*}

\subsection{Quasi-Dynamic Reconstruction}
Now, we explore the potential of our proposed method in a quasi-dynamic environment. We focus on the scenario where the RF Radiance Field of the scene has already been fully measured for time slot $t$, and at time slot $t+1$, some parts of the RF Radiance Field have changed due to environmental dynamics. The received signal power at a target position $P_{Target}$ can be written as:
\begin{align}
\boldsymbol{y}_{t+1,p_{Target}} = \boldsymbol{y}_{t,p_{Target}} + \boldsymbol{e}_{t,p_{Target}} \label{eq: difference equation}
\end{align}
where $\boldsymbol{e}_{t,,p_{Target}}$ denotes the difference of the value between time slot $t$ and $t+1$ for position $p_{Target}$.

In the previous static case, our Local Kernel Estimation and Active Sampling methods helped us focus on the sharp areas of the RF Radiance Field. However, when transitioning from time slot $t$ to $t+1$, the sharp areas might remain unchanged. If we continue focusing on these areas, we may end up observing regions that haven't changed. Therefore, instead of learning the new RF Radiance Field directly, we apply our method to learn the difference between the RF Radiance Fields at time slots $t$ and $t+1$. That is, we obtain $\boldsymbol{y}_{t+1,p_{Target}}$ by learning the difference $\boldsymbol{e}_{t,p_{Target}}$, as shown in (\ref{eq: difference equation}).
By focusing on the differences between the two time slots, our method prioritizes regions with significant changes, which is exactly what is needed to adapt to the evolving RF Radiance Field. Examples will be provided in the next section.

\begin{figure*}[ht!]
\centering
\begin{subfigure}{0.245\textwidth}
\includegraphics[width=0.98\linewidth,height=3.3cm]{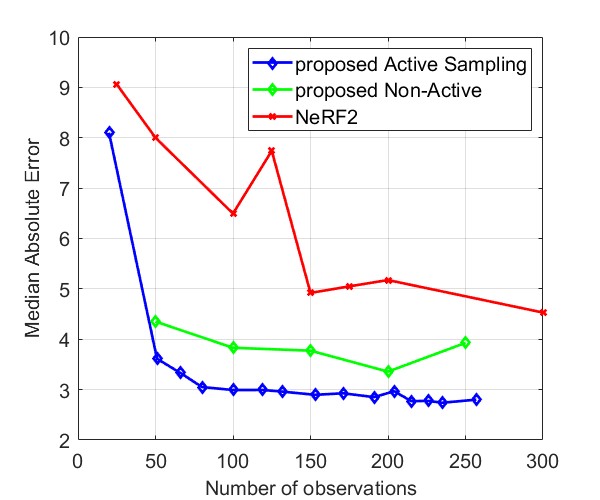} 
\caption{Median AE for Scenario 1}
\end{subfigure}
\begin{subfigure}{0.245\textwidth}
\includegraphics[width=0.98\linewidth,height=3.3cm]{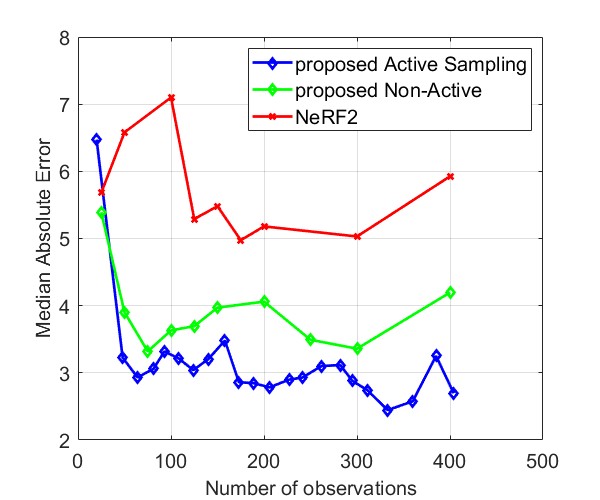}
\caption{Median AE for Scenario 2}
\end{subfigure}
\begin{subfigure}{0.245\textwidth}
\includegraphics[width=0.98\linewidth,height=3.3cm]{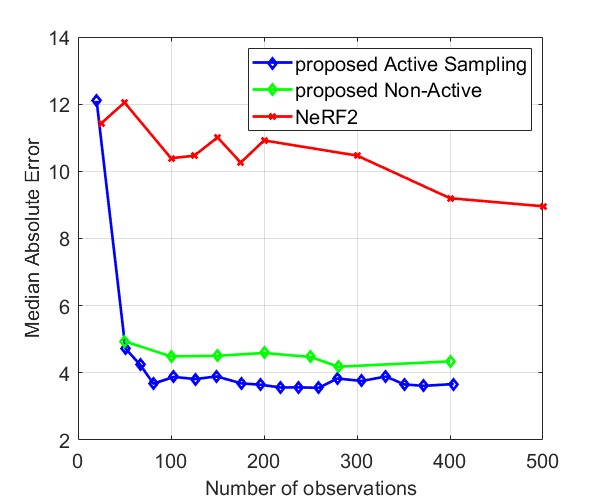}
\caption{Median AE for Scenario 3}
\end{subfigure}
\begin{subfigure}{0.245\textwidth}
\includegraphics[width=0.98\linewidth,height=3.3cm]{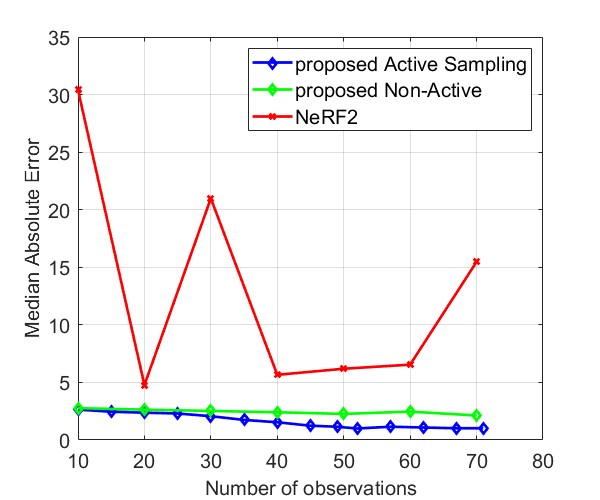}
\caption{Median AE for Scenario 4}
\end{subfigure}

\begin{subfigure}{0.245\textwidth}
\includegraphics[width=0.98\linewidth,height=3.3cm]{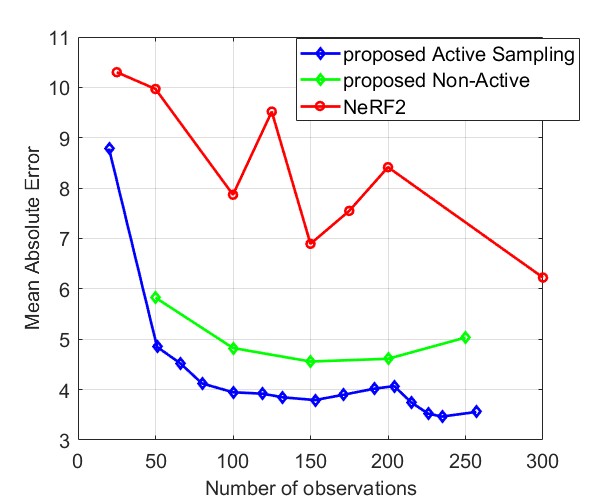} 
\caption{MAE for Scenario 1}
\end{subfigure}
\begin{subfigure}{0.245\textwidth}
\includegraphics[width=0.98\linewidth,height=3.3cm]{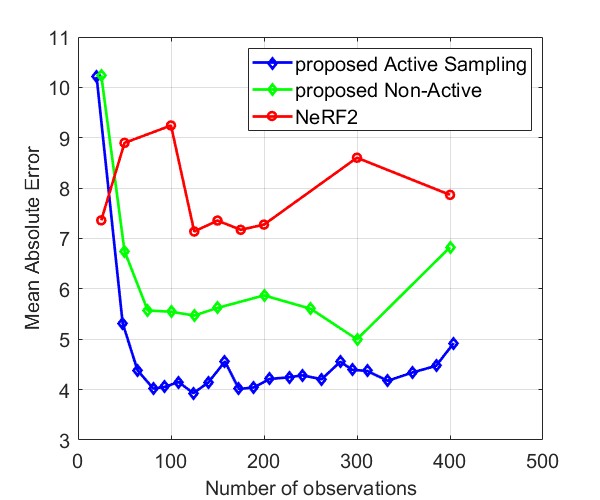}
\caption{MAE for Scenario 2}
\end{subfigure}
\begin{subfigure}{0.245\textwidth}
\includegraphics[width=0.98\linewidth,height=3.3cm]{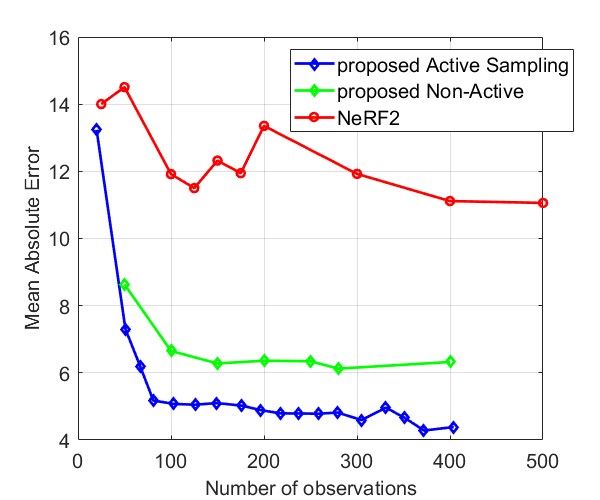}
\caption{MAE for Scenario 3}
\end{subfigure}
\begin{subfigure}{0.245\textwidth}
\includegraphics[width=0.98\linewidth,height=3.3cm]{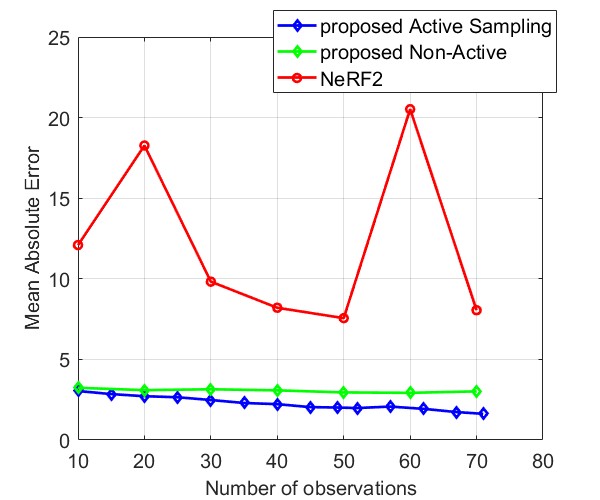}
\caption{MAE for Scenario 4}
\end{subfigure}

\caption{Performance comparison for various scenarios using real data}
\label{fig: Median AE real data}

\end{figure*}

\section{Experiments}
In this section, we present the experimental validation of our proposed method and compare it with NeRF2.

\subsection{Experimental Setup and Data Collection:} We use a set of scenarios consistent with the BLE localization experiments reported in NeRF2: 1) omnidirectional antenna measurements in a single-carrier system, with only the received signal power being recorded; and 2) high operating SNR (negligible noise).   

In the first 3 sets of experiments, we rely solely on the dataset provided in the original NeRF2 paper,  which corresponds to rooms of sizes $10m \times 6m$, $8m \times 4m$, and $8m \times 8m$, respectively. Each scenario is shown in Fig.~\ref{fig: all NeRF2 scenario}, where the red-dots show the reported NeRF2 measurements taken along a trajectory within an indoor scenario with blockers and scatters (Fig. 12 of NeRF2 paper). The 4th dataset was collected in a $10m \times 6m$ meeting room at a local office, following the same procedure, to provide independent evidence. We utilize a Turtlebot4 robot as our mobile platform. The robot carries two key wireless systems: a 60 GHz mmWave system using MikroTik wAP 60G×3 routers, and a 5 GHz WiFi system consisting of an 802.11ac access point and an LG Nexus 5 smartphone using CSIKit to extract received signal strength information (RSSI). 

Additionally, we use the pre-trained NeRF2 model to simulate the received power across the entire room for scenarios 1-3 with a $0.1m$ resolution.

In the Quasi-Dynamic case, we apply a perturbation to the pre-trained NeRF2 model to simulate a new RF Radiance Field. For a more realistic simulation, we use the Nvidia Sionna simulator \cite{hoydis2022sionna} to construct an RF Radiance Field with a $2m \times 2m$ object placed in a $4m \times 4m$ room. The object is then moved by $0.2m$ to a new position. It is important to note that our proposed method does not require prior knowledge of the object's position or the position of the transmitter.

\textbf{Evaluation Metrics:} We evaluate the performance of our algorithm and NeRF2 based on the number of measurements, using the mean/median absolute error between the predicted signal strength of each method and the measured value at the remaining points.

\subsection{Results}
Here, we present the results of the experiment for both static and quasi-dynamic RF Radiance Fields.

\subsubsection{Static RF Radiance Field}
Fig.~\ref{fig: Median AE simulated data} and Fig.~\ref{fig: Median AE real data} compare the performance of our proposed method and NeRF2, using simulated data and real data, respectively. The results for simulated data show that our method outperforms NeRF2, reducing the number of observations by $30\% - 60\%$. For real data, our method also shows better performance, with substantial improvements in both MAE and Median AE. In both cases, our method exhibits much more stable performance, while NeRF2's performance can vary significantly with small changes in the number of samples. This variability arises from the randomness of the initial values of the neural network. With a limited number of samples, the network struggles to converge to a solution with relatively small errors. Furthermore, the results show that active sampling outperforms non-active sampling, as it prioritizes collecting samples from positions with high uncertainty. 

Computation time comparison:
NeRF2 (running on a A6000 server) – 6 hours; our method (running on a 4 core CPU laptop)  – 1 minute total (non-adaptive); 10 seconds per iteration (adaptive)

\subsubsection{Quasi-Dynamic RF Radiance Field}

\begin{figure}
\begin{minipage}[t]{0.45\columnwidth}
  \includegraphics[width=\linewidth]{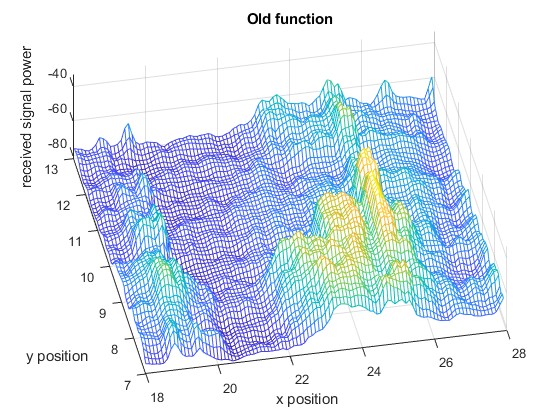}
\caption{Old RF Radiance Field}
\label{fig: old function}
\end{minipage}\hfill 
\begin{minipage}[t]{0.45\columnwidth}
\includegraphics[width=\linewidth]{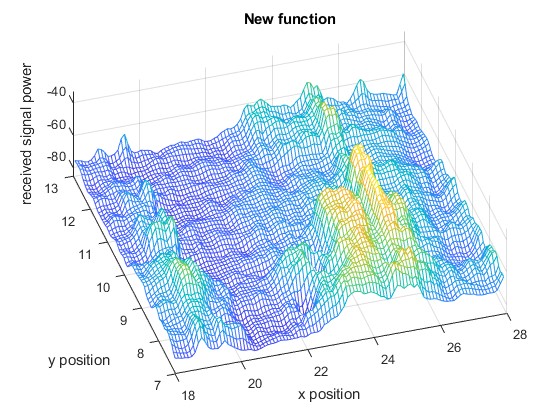}
\caption{New RF Radiance Field}
\label{fig: new function}
\end{minipage}

\begin{minipage}[t]{0.45\columnwidth}
  \includegraphics[width=\linewidth]{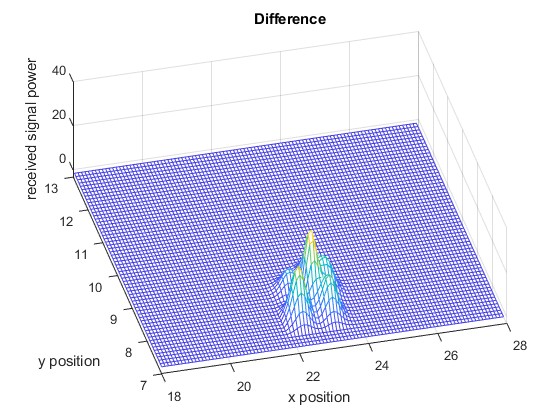}
\caption{The difference of the RF Radiance Field between time slot $t$ and $t+1$}
\label{fig: function change}
\end{minipage}\hfill 
\begin{minipage}[t]{0.45\columnwidth}

\includegraphics[width=\linewidth]{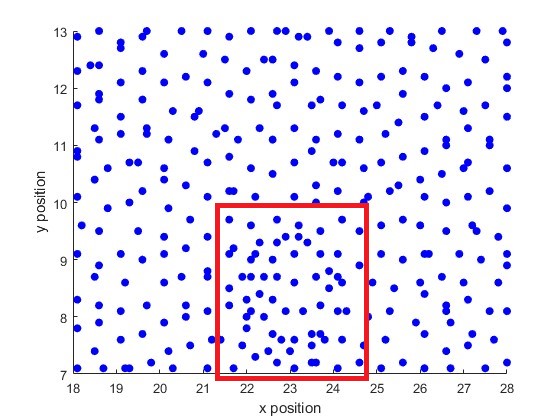}
\caption{Observed positions for learning the difference in the RF Radiance Field}
\label{fig: wakedupSensors Diff}
\end{minipage}

\begin{minipage}[t]{0.45\columnwidth}
  \includegraphics[width=\linewidth,height=3.3cm]{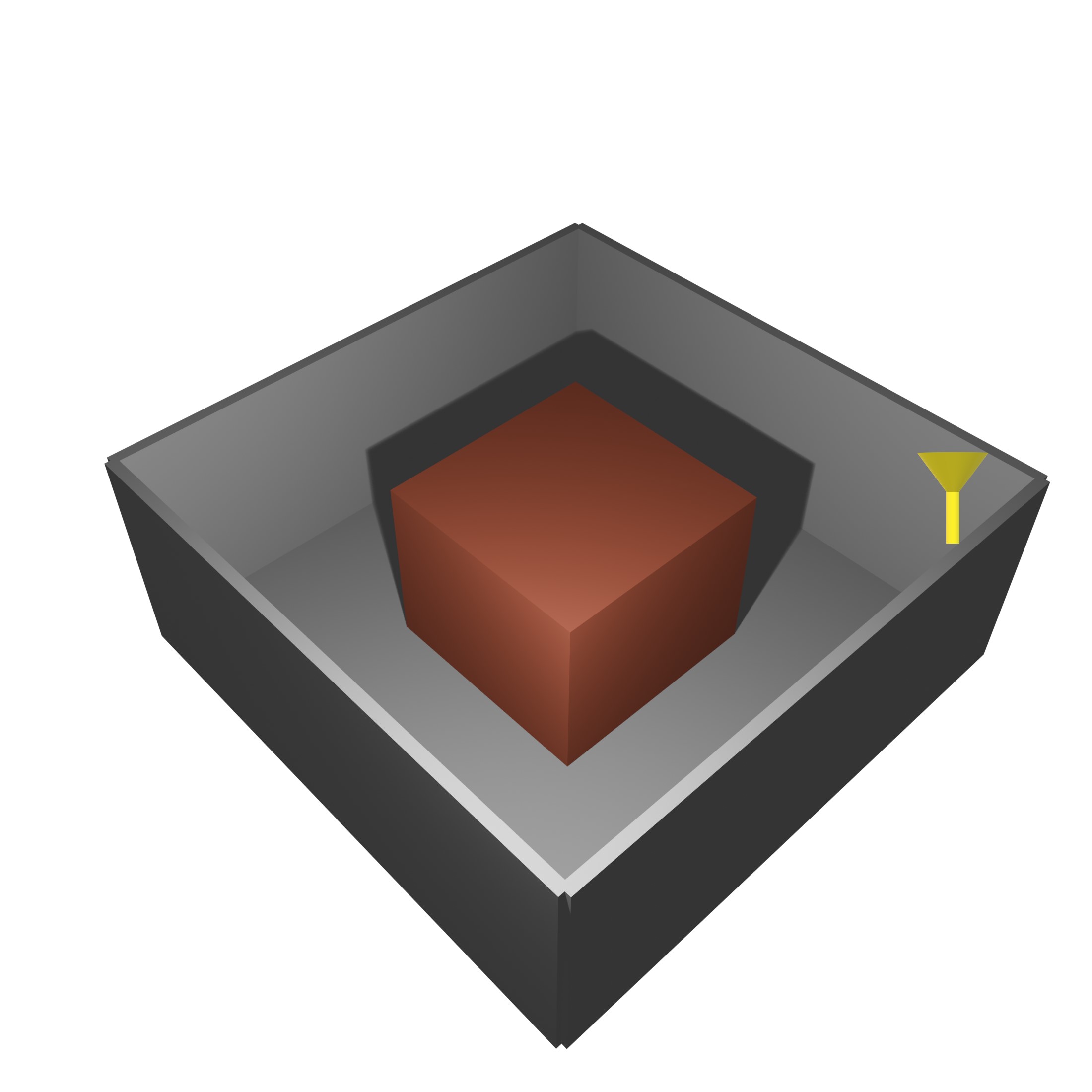}
\caption{Old layout}
\label{fig: old layout}
\end{minipage}\hfill 
\begin{minipage}[t]{0.45\columnwidth}
\includegraphics[width=\linewidth,height=3.3cm]{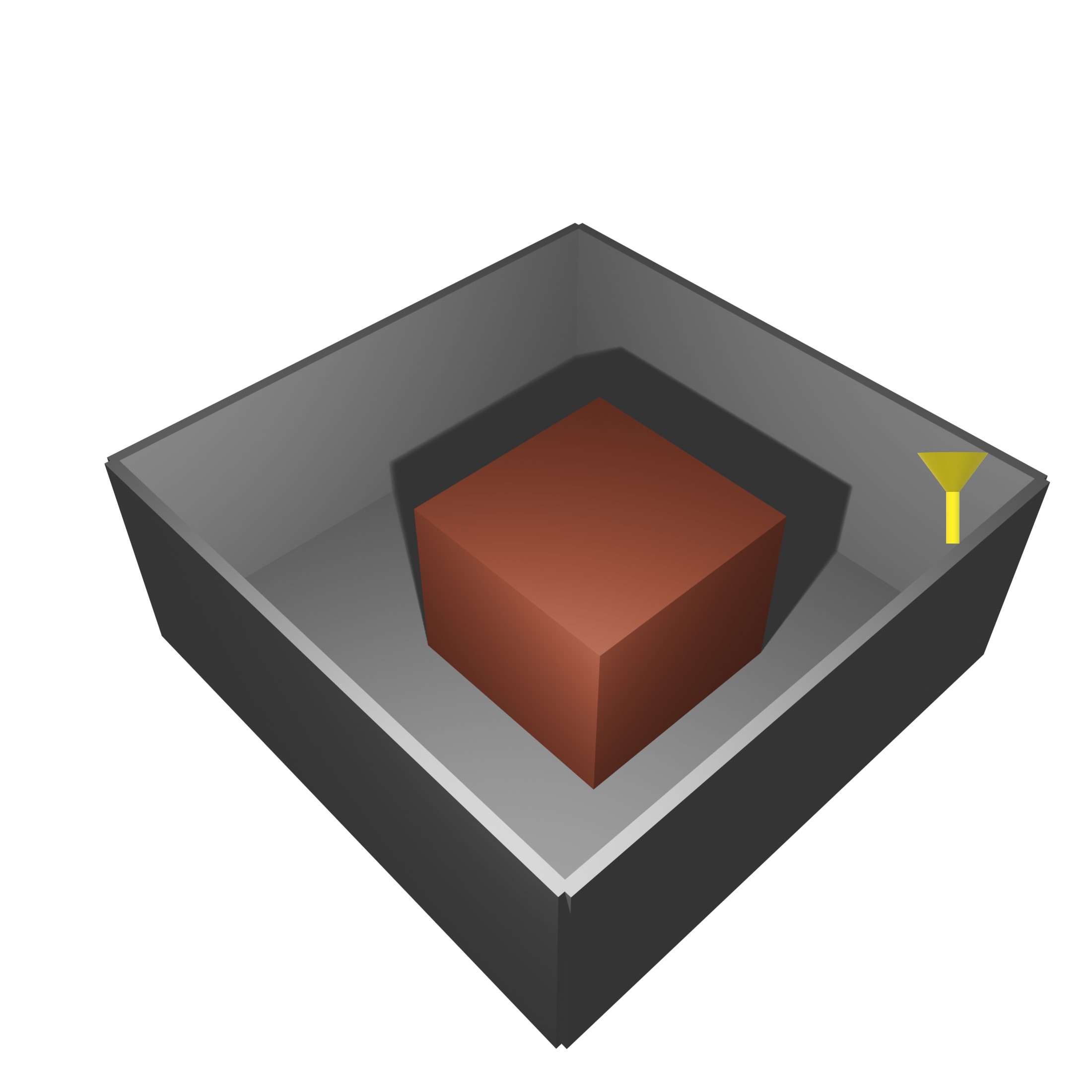}
\caption{New layout}
\label{fig: new layout}
\end{minipage}

\begin{minipage}[t]{0.45\columnwidth}
  \includegraphics[width=\linewidth]{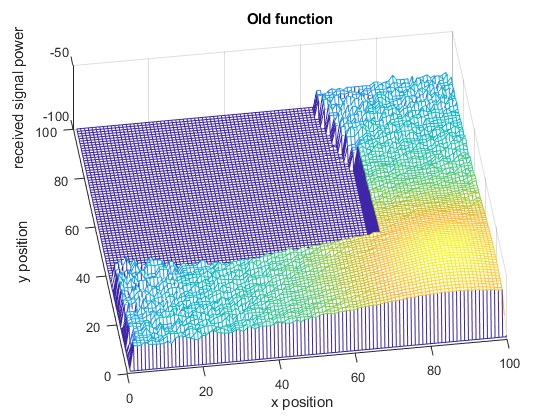}
\caption{Old RF Radiance Field}
\label{fig: old function 2}
\end{minipage}\hfill 
\begin{minipage}[t]{0.45\columnwidth}
\includegraphics[width=\linewidth]{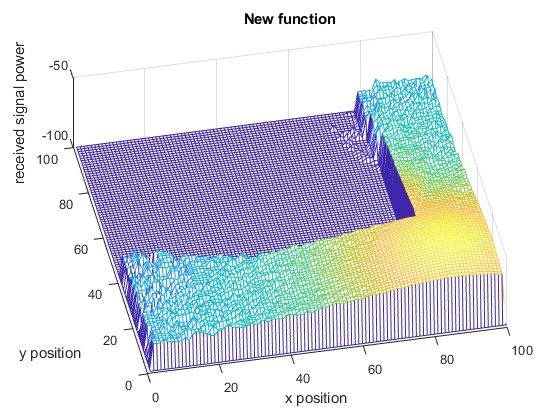}
\caption{New RF Radiance Field}
\label{fig: new function 2}
\end{minipage}

\begin{minipage}[t]{0.45\columnwidth}
  \includegraphics[width=\linewidth]{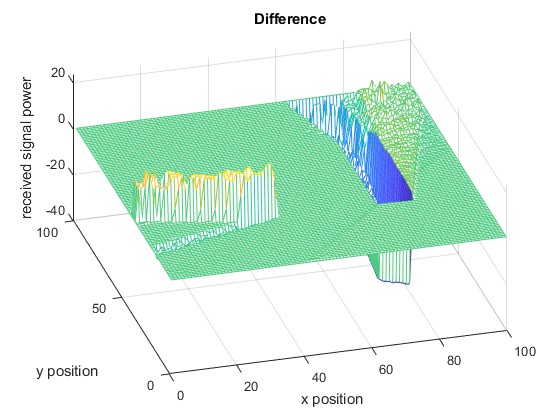}
\caption{The difference of the RF Radiance Field between time slot $t$ and $t+1$}
\label{fig: function change 2}
\end{minipage}\hfill 
\begin{minipage}[t]{0.45\columnwidth}

\includegraphics[width=\linewidth]{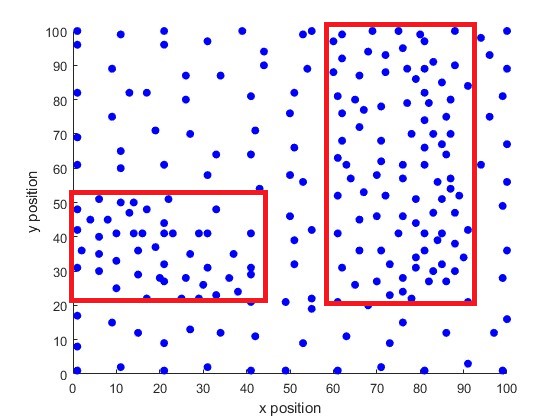}
\caption{Observed positions for learning the difference in the RF Radiance Field}
\label{fig: wakedupSensors Diff 2}
\end{minipage}
\vspace{-10pt}
\end{figure}

\begin{figure}
\begin{minipage}[t]{0.45\columnwidth}
  \includegraphics[width=\linewidth]{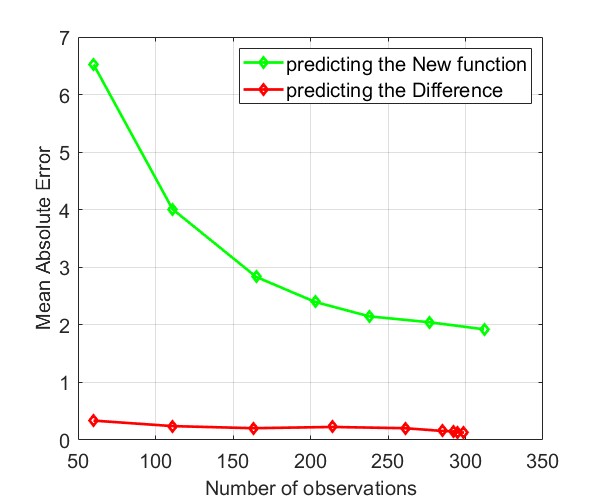}
\caption{MAE Comparison (from NeRF2)}
\label{fig: MAE comparison Dynamic}
\end{minipage}\hfill 
\begin{minipage}[t]{0.45\columnwidth}

\includegraphics[width=\linewidth]{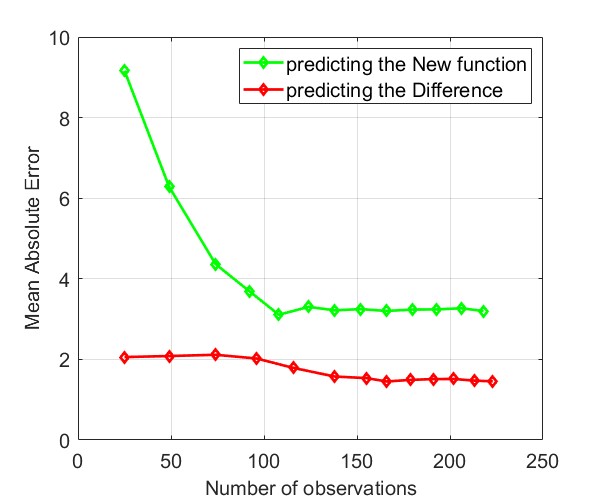}
\caption{MAE Comparison (from Nvidia Sionna)}
\label{fig: MAE comparison Dynamic 2}
\end{minipage}
\vspace{-10pt}
\end{figure}


Fig.~\ref{fig: old function} and Fig.~\ref{fig: new function} show the RF Radiance Fields generated by the pre-trained NeRF2 model at times $t$ and $t+1$, with their difference in Fig.~\ref{fig: function change}. Learning the new RF Radiance Field can lead to excessive sampling of areas that remain unchanged and sharp. In contrast, our method focuses on learning the differences between the fields, prioritizing the areas that have changed. As shown in Fig.~\ref{fig: wakedupSensors Diff}, the high-density area (marked by the red square) in the collected observations, used to learn the difference, corresponds to the region where the RF Radiance Field has changed. Additionally, as seen in Fig.~\ref{fig: MAE comparison Dynamic}, learning the difference allows us to achieve better performance with fewer samples.

Fig.~\ref{fig: old layout} and Fig.~\ref{fig: new layout} show the layout of the scene with a signal source placed at the right corner for two different time slots, while Fig.~\ref{fig: old function 2} and  Fig.~\ref{fig: new function 2} illustrate the corresponding RF Radiance Fields, generated by the NVIDIA Sionna model.
 Fig.~\ref{fig: function change 2}, show the differences between them. We can observe that, since the object has been moved (shifted by $0.2m$), part of the area experiences a signal power drop, while areas that were previously blocked now receive a stronger signal. Similar to the previous example,  Fig.~\ref{fig: wakedupSensors Diff 2} shows that the high-density area (marked with red squares) of the collected samples corresponds to the region where the RF Radiance Field has changed. The MAE comparison is shown in Fig.~\ref{fig: MAE comparison Dynamic 2}.


It is shown that our approach allows for updating the RF Radiance Field by learning the differences with fewer samples, compared to directly learning the entire new function. This makes it more efficient and practical for real-world applications, particularly in computer vision tasks like 3D scene reconstruction, object tracking, and visual SLAM (Simultaneous Localization and Mapping). By reducing the sample requirements, our method facilitates quicker updates in dynamic environments, a key advantage for real-time applications such as augmented reality (AR) and autonomous navigation, where scenes continuously evolve and need to be reconstructed on the fly.




 
\section{Conclusions}





In this work, we propose a training-free method for predicting received signal power in RF Radiance Fields. Our approach generates predictions quickly at any location within the scene and includes an uncertainty model. By utilizing local kernel estimation, it offers improved computational efficiency over traditional Gaussian prediction methods. Simulation results show that fewer samples are needed to achieve performance comparable to NeRF2. Additionally, we further reduce the number of required observations by incorporating active sampling, which selects the most informative data points. Importantly, our method is well-suited for dynamic scenes, enabling near-instantaneous reconstruction as soon as new observations are available. This real-time adaptability is essential for computer vision applications in dynamic and constantly evolving environments.
\newpage
{
    \small
    \bibliographystyle{ieeenat_fullname}
    \bibliography{main}

\begin{thebibliography}{18}
\providecommand{\natexlab}[1]{#1}
\providecommand{\url}[1]{\texttt{#1}}
\expandafter\ifx\csname urlstyle\endcsname\relax
  \providecommand{\doi}[1]{doi: #1}\else
  \providecommand{\doi}{doi: \begingroup \urlstyle{rm}\Url}\fi

\bibitem[Chan et~al.(2019)Chan, Saito, Li, Zhao, Xiao, Sugimoto, Sakashita, and Komura]{chan2021pifu}
Shunsuke Chan, Shunsuke Saito, Lingyu Li, Hao Zhao, Shigeo Xiao, Kenji Sugimoto, Shohei Sakashita, and Taku Komura.
\newblock Pifu: Pixel-aligned implicit function for high-resolution clothed human digitization.
\newblock \emph{Proceedings of the IEEE/CVF International Conference on Computer Vision}, 2019.

\bibitem[Chen et~al.(2024)Chen, Feng, Sun, Qian, and Zhang]{chen2024rfcanvas}
Xingyu Chen, Zihao Feng, Ke Sun, Kun Qian, and Xinyu Zhang.
\newblock Rfcanvas: Modeling rf channel by fusing visual priors and few-shot rf measurements.
\newblock In \emph{Proceedings of the 22nd ACM Conference on Embedded Networked Sensor Systems}, pages 464--477, 2024.

\bibitem[Hoydis et~al.(2022)Hoydis, Cammerer, Aoudia, Vem, Binder, Marcus, and Keller]{hoydis2022sionna}
Jakob Hoydis, Sebastian Cammerer, Fay{\c{c}}al~Ait Aoudia, Avinash Vem, Nikolaus Binder, Guillermo Marcus, and Alexander Keller.
\newblock Sionna: An open-source library for next-generation physical layer research.
\newblock \emph{arXiv preprint arXiv:2203.11854}, 2022.

\bibitem[Hu et~al.(2023)Hu, Zheng, Chen, Wang, and Luo]{hu2023muse}
Jingzhi Hu, Tianyue Zheng, Zhe Chen, Hongbo Wang, and Jun Luo.
\newblock Muse-fi: Contactless muti-person sensing exploiting near-field wi-fi channel variation.
\newblock In \emph{Proceedings of the ACM MobiCom}, 2023.

\bibitem[Kerbl et~al.(2023)Kerbl, Kopanas, Leimk{\"u}hler, and Drettakis]{kerbl20233d}
Bernhard Kerbl, Georgios Kopanas, Thomas Leimk{\"u}hler, and George Drettakis.
\newblock 3d gaussian splatting for real-time radiance field rendering.
\newblock \emph{ACM Trans. Graph.}, 42\penalty0 (4):\penalty0 139--1, 2023.

\bibitem[Mildenhall et~al.(2020)Mildenhall, Srinivasan, Tancik, Barron, Ramamoorthi, and Ng]{mildenhall2020nerf}
Ben Mildenhall, Pratul~P Srinivasan, Matthew Tancik, Jonathan~T Barron, Ravi Ramamoorthi, and Ren Ng.
\newblock Nerf: Representing scenes as neural radiance fields for view synthesis.
\newblock In \emph{European Conference on Computer Vision}, 2020.

\bibitem[Mildenhall et~al.(2021)Mildenhall, Srinivasan, Tancik, Barron, Ramamoorthi, and Ng]{mildenhall2021nerf}
Ben Mildenhall, Pratul~P Srinivasan, Matthew Tancik, Jonathan~T Barron, Ravi Ramamoorthi, and Ren Ng.
\newblock Nerf: Representing scenes as neural radiance fields for view synthesis.
\newblock \emph{Communications of the ACM}, 65\penalty0 (1):\penalty0 99--106, 2021.

\bibitem[Orekondy et~al.(2023)Orekondy, Kumar, Kadambi, Ye, Soriaga, and Behboodi]{orekondy2023winert}
Tribhuvanesh Orekondy, Pratik Kumar, Shreya Kadambi, Hao Ye, Joseph Soriaga, and Arash Behboodi.
\newblock Winert: Towards neural ray tracing for wireless channel modelling and differentiable simulations.
\newblock In \emph{The Eleventh International Conference on Learning Representations}, 2023.

\bibitem[Park et~al.(2019)Park, Florence, Straub, Newcombe, and Lovegrove]{park2019deepsdf}
Jeong~Joon Park, Peter Florence, Julian Straub, Richard Newcombe, and Steven Lovegrove.
\newblock Deepsdf: Learning continuous signed distance functions for shape representation.
\newblock \emph{Proceedings of the IEEE/CVF Conference on Computer Vision and Pattern Recognition}, 2019.

\bibitem[Sitzmann et~al.(2020)Sitzmann, Martel, Bergman, Lindell, and Wetzstein]{sitzmann2020implicit}
Vincent Sitzmann, Julien~N.P. Martel, Alexander~W. Bergman, David~B. Lindell, and Gordon Wetzstein.
\newblock Implicit neural representations with periodic activation functions.
\newblock In \emph{Advances in Neural Information Processing Systems}, 2020.

\bibitem[Tewari et~al.(2020)Tewari, Fried, Thies, Sitzmann, Lombardi, Sunkavalli, Martin-Brualla, Simon, Saragih, Nießner, et~al.]{tewari2020state}
Ayush Tewari, Ohad Fried, Justus Thies, Vincent Sitzmann, Stephen Lombardi, Kalyan Sunkavalli, Ricardo Martin-Brualla, Tomas Simon, Jason Saragih, Matthias Nießner, et~al.
\newblock State of the art on neural rendering.
\newblock \emph{Computer Graphics Forum}, 39\penalty0 (2), 2020.

\bibitem[Vakalis et~al.(2019)Vakalis, Gong, and Nanzer]{vakalis2019imaging}
Stavros Vakalis, Liang Gong, and Jeffrey~A Nanzer.
\newblock Imaging with wifi.
\newblock \emph{IEEE Access}, 7:\penalty0 28616--28624, 2019.

\bibitem[Williams and Rasmussen(2006)]{williams2006gaussian}
Christopher~KI Williams and Carl~Edward Rasmussen.
\newblock \emph{Gaussian processes for machine learning}.
\newblock MIT press Cambridge, MA, 2006.

\bibitem[Woodford et~al.(2022)Woodford, Zhang, Chai, and Sundaresan]{woodford2022mosaic}
Timothy Woodford, Xinyu Zhang, Eugene Chai, and Karthikeyan Sundaresan.
\newblock Mosaic: Leveraging diverse reflector geometries for omnidirectional around-corner automotive radar.
\newblock In \emph{Proceedings of the 20th Annual International Conference on Mobile Systems, Applications and Services (MobiSys)}, 2022.

\bibitem[Xie et~al.(2021)Xie, Takikawa, Saito, Litany, Yan, Khan, Tombari, Tompkin, Sitzmann, and Sridhar]{xie2021neural}
Yiheng Xie, Towaki Takikawa, Shunsuke Saito, Or Litany, Shiqin Yan, Numair Khan, Federico Tombari, James Tompkin, Vincent Sitzmann, and Srinath Sridhar.
\newblock Neural fields in visual computing and beyond.
\newblock In \emph{Computer Graphics Forum}, 2021.

\bibitem[Yariv et~al.(2020)Yariv, Kasten, Moran, Galun, Atzmon, Ronen, and Lipman]{yariv2020multiview}
Lior Yariv, Yoni Kasten, Dror Moran, Meirav Galun, Matan Atzmon, Basri Ronen, and Yaron Lipman.
\newblock Multiview neural surface reconstruction by disentangling geometry and appearance.
\newblock In \emph{Advances in Neural Information Processing Systems}, 2020.

\bibitem[Zhao et~al.(2022)Zhao, Huang, and Shen]{zhao2022high}
Hanying Zhao, Mingtao Huang, and Yuan Shen.
\newblock High-accuracy localization in multipath environments via spatio-temporal feature tensorization.
\newblock \emph{IEEE Transactions on Wireless Communications}, 21\penalty0 (12):\penalty0 10576--10591, 2022.

\bibitem[Zhao et~al.(2023)Zhao, An, Pan, and Yang]{zhao2023nerf2}
Xiaopeng Zhao, Zhenlin An, Qingrui Pan, and Lei Yang.
\newblock Nerf2: Neural radio-frequency radiance fields.
\newblock In \emph{Proceedings of the 29th Annual International Conference on Mobile Computing and Networking}, pages 1--15, 2023.

\end{thebibliography}

}


\end{document}